\documentclass[preprint,a4paper]{aastex631}

\usepackage[utf8]{inputenc}
\usepackage{textgreek}
\usepackage{amsmath,amssymb,bm} 
\usepackage{booktabs}
\usepackage{natbib}
\usepackage{graphicx}
\usepackage{upgreek}

\shorttitle{Filament Fragment Association}
\shortauthors{Zhu et al.}

\begin{document}

\title{Graph Learning with Multiplicative Aggregation for Solar Filament Fragment Association Modeling in Full-disk H$\alpha$ Images}

\correspondingauthor{GaoFei Zhu}
\email{gfzhu@xhu.edu.cn}

\author[0000-0002-0405-7018]{GaoFei Zhu}
\affiliation{School of Computer and Software Engineering, Xihua University, Chengdu 610039, P. R. China}
\affiliation{State Key Laboratory of Solar Activity and Space Weather, National Astronomical Observatories, Chinese Academy of Sciences, Beijing 100101, P. R. China}

\author{GangHua Lin}
\affiliation{State Key Laboratory of Solar Activity and Space Weather, National Astronomical Observatories, Chinese Academy of Sciences, Beijing 100101, P. R. China}

\author[0000-0003-1675-1995]{Xiao Yang}
\affiliation{State Key Laboratory of Solar Activity and Space Weather, National Astronomical Observatories, Chinese Academy of Sciences, Beijing 100101, P. R. China}

\author[0000-0002-1396-7603]{Suo Liu}
\affiliation{State Key Laboratory of Solar Activity and Space Weather, National Astronomical Observatories, Chinese Academy of Sciences, Beijing 100101, P. R. China}

\author[0000-0002-7694-2454]{Yu Liu}
\affiliation{School of Physical Science and Technology, Southwest Jiaotong University, Chengdu 611756, P. R. China}

\author{ZhiSheng Gao}
\affiliation{School of Computer and Software Engineering, Xihua University, Chengdu 610039, P. R. China}

\author{Cheng Zeng}
\affiliation{School of Computer and Software Engineering, Xihua University, Chengdu 610039, P. R. China}

\begin{abstract}
Solar filaments are prominent features in H$\alpha$ observations. During their evolution, they often exhibit complex morphological and topological changes, which, due to line-of-sight projection effects, appear as fragmentation and merging. These phenomena make it difficult to reliably track solar filaments using automated methods, as the same structure may appear different over time. To address this issue, we formulate filament fragment association as a graph-based learning problem, where individual fragments are represented as nodes and candidate pairwise relationships are modeled as edges to infer whether fragments belong to the same filament. Furthermore, we propose a novel graph neural network (GNN) that incorporates multiplicative aggregation layers to enhance feature aggregation in the graph, replacing conventional additive aggregation with a multiplicative mechanism to better capture information between fragments. Experimental results demonstrate that the proposed method achieves reliable performance in filament fragment association, validating the effectiveness of the graph-based formulation and the proposed multiplicative aggregation mechanism.

\end{abstract}


\section{Introduction} 
\label{sec:intro}

Solar filaments, typically observed as elongated dark structures in H$\alpha$ images, are cool and dense plasmas that are often suspended in the solar corona by magnetic fields \citep{chen2020some}. Solar filaments are closely associated with eruptive phenomena such as coronal mass ejections (CMEs) \citep{gilbert2000active,gopalswamy2003prominence,jing2004relation,chen2008initiation,sinha2019solar}. Therefore, analyzing their morphological structures and evolutionary processes is of great importance for understanding solar activity and improving space weather forecasting. However, when observed in H$\alpha$ images from the line-of-sight perspective, solar filaments often exhibit complex evolution, including fragmentation \citep{chen2023observations}, merging \citep{jiang2014interaction,joshi2014formation}, eruption \citep{schmieder2013solar,zou2019statistical,hou2023partial}, and disappearance, which leads to discontinuous and fragmented structures on the solar disk. Previous studies have demonstrated that magnetic processes, such as magnetic reconnection, play a key role in driving the morphological and topological evolution of solar filaments \citep{yang2016rapid, sinha2019solar,joshi2023interaction,liu2024filament}. This implies that the morphological evolution of solar filaments, including processes such as fragmentation and merging, is intrinsically linked to underlying physical processes. Furthermore, long-term statistical analyses of solar filament properties offer valuable insights into their physical properties and evolutionary behavior, including length, spatial distribution, orientation, latitudinal migration, and chirality. 

The automatic segmentation of solar filaments in H$\alpha$ images has been extensively studied and is an important prerequisite for subsequent research, such as evolutionary tracking and physical properties extraction \citep{hao2015statistical}. In earlier studies, intensity thresholding was widely used to separate filaments from the background, including global thresholding, local thresholding, and region growing techniques \citep{gao2002development,shih2003automatic,fuller2005filament,hao2013developing,hao2015statistical}. These methods are simple and efficient, but often lack robustness in complex scenarios. Subsequently, traditional machine learning methods (such as support vector machines and neural networks) were introduced for filament detection \citep{zharkova2005filament,labrosse2010automatic}. However, these methods are typically applicable to local regions or specific scenes (such as limb prominences), exhibiting limited generalization ability. In recent years, deep learning methods have been widely applied to filament segmentation. Models such as Mask R-CNN and U-Net have significantly improved detection performance by learning features from data. Among them, U-Net-based methods are widely adopted due to their simple structure and excellent segmentation results \citep{zhu2019solar}. Subsequent studies improved performance by refining network structures and introducing attention mechanisms and multi-scale features, leading to improved accuracy and robustness in filament detection and segmentation in H$\alpha$ observations \citep{liu2021solar,mackovjak2021scss,guo2022solar,diercke2024universal,zhu2025flat}.

Filaments may split into multiple fragments during their evolution, which makes subsequent tracking and physical parameter extraction challenging. In particular, during the tracking process, determining which filament fragments originate from the same event is crucial for understanding the dynamical evolution of solar filaments. \cite{gao2002development} and \cite{shih2003automatic} use a fixed distance threshold to logically merge filament segments. The advantage lies in its computational simplicity, but its general applicability is clearly limited. To improve this, \cite{bernasconi2005advanced} further incorporate geometric constraints. In addition to spatial distance, they consider the orientation of filament segments, merging segments that are not only close to each other but also aligned in a consistent direction. Such strategies improve the reliability of merging fragmented filaments to some extent, but they still rely on handcrafted rules and may fail in more complex scenarios. \cite{joshi2010automated} and \cite{hao2013developing} also employ a similar strategy, logically merging filament fragments into a single filament. In recent years, deep learning methods have been explored to address the filament fragment association problem. \cite{guo2022solar} proposed a method based on the instance segmentation model CondInst to associate filament fragments. Transforming the filament fragment association problem into an instance segmentation task is an appealing modeling strategy, enabling pixel-level segmentation and fragment association to be handled within a single model. Unlike conventional instance segmentation tasks where each instance usually corresponds to an independent physical object, filament fragments may represent disconnected parts of the same physical filament due to observational limitations or intrinsic evolution. However, instance segmentation models are typically more complex, as they jointly optimize detection and segmentation. As a result, they may struggle to achieve pixel-level accuracy comparable to dedicated semantic segmentation models (e.g., U-Net), especially when training data are limited or filament structures are irregular. Furthermore, from a modeling perspective, treating multiple fragments from the same physical filament as separate instances may conflict with the assumption of instance segmentation that each instance represents an independent object. Therefore, the correspondence between predicted instances and physical filaments becomes ambiguous, since multiple predicted instances may originate from a single physical filament. Such a representation, which treats fragments of a single physical filament as multiple pseudo-instances, may introduce semantic ambiguity at the label level. Nevertheless, it still provides a meaningful perspective for modeling the filament fragment association problem.

The solar filament fragment association problem is inherently ambiguous, as fragments from the same filament may be spatially separated, while fragments from different filaments may appear closely located or structurally similar. As a result, spatial proximity and orientation are not sufficient for reliable association. To mitigate this ambiguity, more comprehensive features that encode global structure and interfragment relationships must be considered. We propose a graph-based learning framework that explicitly models both local geometry and global relational context. In this framework, filament fragments are represented as nodes with multiscale features capturing geometric, photometric, and structural properties, while edges encode spatial proximity and pairwise relationships. The association problem is then formulated as a link prediction task, allowing the model to infer connectivity between fragments in a data-driven manner.

Our paper is organized as follows. Section~\ref{sec:methods} describes the proposed graph-based learning framework, including node feature extraction, graph construction, and solar filament fragment association prediction. Section~\ref{sec:exp} presents data description, experimental setup, and quantitative results, along with comparisons to baseline methods. Finally, Section~\ref{sec:dis_con} summarizes the main results and discusses the limitations of the proposed method, along with potential directions for future improvement.

\section{Methods} 
\label{sec:methods}

\subsection{Filament Fragment Association Framework}

\begin{figure}[!htbp]
\centering
\plotone{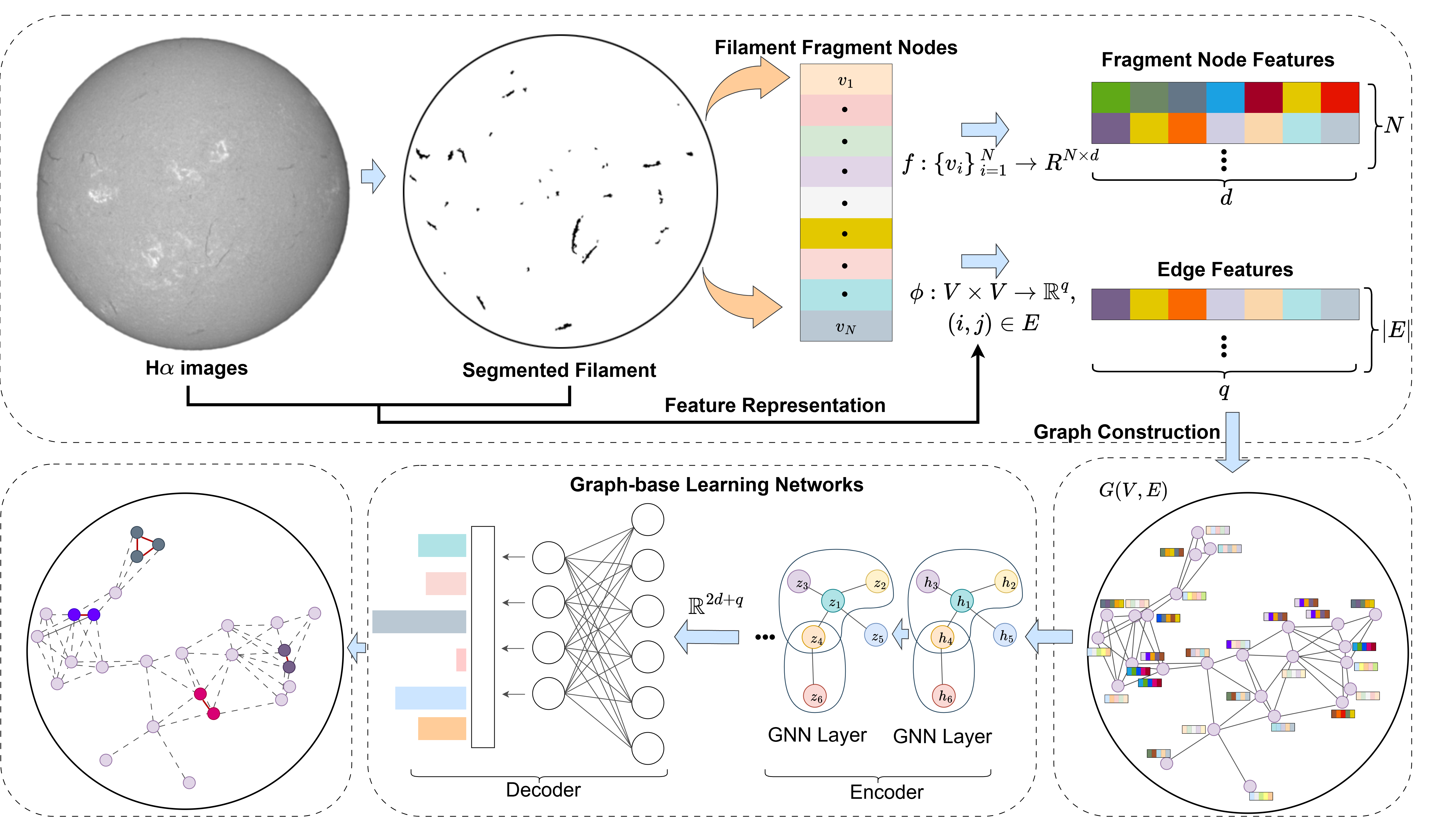}
\caption{Overview of the proposed graph-based learning framework for solar filament fragment association.}
\label{fig:framework}
\end{figure}

Solar filament fragment association is formulated as a graph-based relational learning task. As shown in Figure~\ref{fig:framework}, we construct a graph representation in which segmented fragments are treated as nodes and their pairwise relationships are explicitly modeled. Starting from H$\alpha$ solar images, a segmentation procedure is first applied to extract filament structures, producing a set of disconnected filament fragments. These fragments are typically spatially separated due to observational limitations or intrinsic structural complexity, which makes it necessary to determine whether disconnected fragments belong to the same filament. Instead of relying primarily on spatial distance thresholds and orientation, we represent filament fragments as graph nodes, enabling comprehensive inference of their associations by jointly considering node attributes and pairwise relationships.

For each filament fragment, we construct a set of node features to encode its intrinsic properties, including geometric scale, spatial location, morphological structure, and photometric characteristics. Formally, this process is defined by a mapping 
$f:\{v_i\}_{i=1}^{N} \rightarrow \mathbb{R}^{N \times d}$, 
which assigns each node a $d$-dimensional feature vector. These features provide complementary information by capturing multiple aspects of each fragment, allowing fragments with similar appearances but different physical origins to be distinguished. In parallel, we explicitly model pairwise relationships between fragments through edge features that capture spatial distance and structural similarity. This is achieved via a mapping $\phi: V \times V \rightarrow \mathbb{R}^{q}$, where \(i\) and \(j\) denote fragment nodes and \((i,j)\in E\) indicates their interaction. This dual representation enables the model to jointly leverage information from individual fragments and their pairwise relationships, which is essential for reliable association.

A graph is constructed for each image, where nodes correspond to filament fragments and edges represent potential associations between fragment pairs. The resulting graph $G^{(n)} = (V^{(n)}, E^{(n)})$ serves as input to a graph neural network (GNN). Within this framework, node representations are iteratively updated via message passing, where information is propagated and aggregated from neighboring nodes. This process leverages higher-order structural information to enhance node representations. Building upon the learned node embeddings, a decoder (or link prediction head) is used to estimate the probability that two fragments belong to the same physical filament. In this way, the fragment association problem is naturally formulated as an edge-level prediction task, where the model distinguishes true connections from spurious ones based on both local features and global graph structure.

In summary, the proposed framework provides a unified and flexible approach to the association of filamentary fragments. By combining explicit feature engineering with graph-based relational modeling, it effectively captures the complex spatial and structural dependencies between fragments, offering new insights into the association problem of solar filament fragments.

\subsection{Graph Construction}
Given a set of H\(\alpha\) solar images \(\mathcal{I} = \left\{ I^{(1)}, I^{(2)}, \cdots, I^{(M)} \right\}\), filament fragment regions are extracted from each image using off-the-shelf segmentation methods. For the \(n\)th image \(I^{(n)},\) a set of connected components corresponding to filament fragments is obtained:
\[
\Omega^{(n)}=\left\{ P_{1}^{(n)},{P_2}^{(n)},\cdots,P_{N_n}^{(n)} \right\},
\]
where \(P_{i}^{(n)}\) denotes the \(i\)th filament fragment in image \(I^{(n)},\) and \(N_n\) is the total number of fragments in the image. Each fragment is treated as a graph node, and thus each image is represented as an individual graph:
\[
G^{(n)}=(V^{(n)},E^{(n)}),
\]
where the node set is defined as:
\[
V^{(n)}=\left\{ v_{1}^{(n)},{v_2}^{(n)},\cdots,v_{N_n}^{(n)} \right\},
\]
with a one-to-one correspondence $v_i^{(n)}\leftrightarrow P_i^{(n)}$.

\subsection{Node Feature Representation} 
To characterize each node \( v_i^{(n)} \), we define a feature extraction function:
\[
\bm{F}_i^{(n)} = \psi\!\big(P_i^{(n)}, I^{(n)}\big), \quad 
\bm{F}_i^{(n)} \in \mathbb{R}^d,\; i = 1, \cdots, N_n,
\]
which encodes the intrinsic properties of each filament fragment by aggregating geometric, spatial, morphological, and photometric cues, resulting in a 13-dimensional feature vector (i.e., \( d = 13 \)) for each node. Such a design enables the model to jointly capture the shape, location, structural continuity, and intensity consistency of fragments, all of which are critical for reliable fragment association.

Geometric features describe the size and extent of each fragment, which are essential for distinguishing large-scale structures from small, noisy regions. Specifically, we compute
\[
\bm{F}_{i,\mathrm{g}}^{(n)} = 
[\log A_i,\; \log \mathrm{Peri}_i,\; \log w_i,\; \log h_i,\; r_i],
\]
where $A_i = |P_i^{(n)}|$ denotes the area of the $i$th filament fragment (i.e., the number of pixels in \( P_i^{(n)} \)), \( \mathrm{Peri}_i \) is the perimeter, \( w_i \) and \( h_i \) represent the width and height of its bounding box, respectively, and \( r_i = \frac{w_i}{h_i} \) denotes the aspect ratio. Furthermore, \( \log(\cdot) \) is applied to scale-related attributes (e.g., area, perimeter, width, and height) to reduce dynamic range and improve numerical stability.

To incorporate spatial context, we further define the centroid-based spatial feature as
\[
\bm{F}_{i,\mathrm{s}}^{(n)} = 
[\tilde{c}_i^x,\; \tilde{c}_i^y],
\]
where \( (c_i^x, c_i^y) \) denotes the centroid coordinates of the \( i \)-th filament fragment, \( W \) and \( H \) are the width and height of the input H$\alpha$ solar image, respectively, and \( \tilde{c}_i^x = \frac{c_i^x}{W} \), \( \tilde{c}_i^y = \frac{c_i^y}{H} \) are the normalized centroid coordinates.

The structural characteristics of filament fragments are defined as
\[
\bm{F}_{i,\mathrm{t}}^{(n)} = 
[\log L_i,\; \epsilon_i,\; \sin\theta_i,\; \cos\theta_i],
\]
where \(L_i\) denotes the skeleton length of the \(i\)th filament fragment, \(\epsilon_i\) is the eccentricity, and \(\theta_i\) represents the orientation angle of the fragment.

The intensity-based features of filament fragments are given by
\[
\mathbf{F}_{i,\mathrm{a}}^{(n)} = 
[\hat{\mu}_i,\; \hat{\sigma}_i],
\]
where \( \mu_i \) and \( \sigma_i \) denote the mean intensity and intensity variance of the \( i \)th filament fragment, respectively, computed as
\[
\mu_i = \frac{1}{\left| P_i^{(n)} \right|} \sum_{(x,y)\in P_i^{(n)}} I^{(n)}(x,y), \quad
\sigma_i = \frac{1}{\left| P_i^{(n)} \right|} \sum_{(x,y)\in P_i^{(n)}} \left( I^{(n)}(x,y) - \mu_i \right)^2,
\]
where \( P_i^{(n)} \) denotes the pixel set of the \( i \)th fragment and \( I^{(n)}(x,y) \) is the intensity at location \( (x,y) \). To mitigate the effect of global intensity variations, the features are further normalized:
\[
\hat{\mu}_i = \frac{\mu_i - \mu_g}{\sigma_g}, \quad
\hat{\sigma}_i = \frac{\sigma_i}{\sigma_g},
\]
where \( \mu_g \) and \( \sigma_g \) denote the global mean and standard deviation of the image, respectively.

The final node representation is constructed by concatenating geometric, spatial, structural, and intensity features, which provide complementary descriptions of filament fragments from multiple perspectives. Specifically,
\[
\bm{F}_i^{(n)} =
\big[
\bm{F}_{i,\mathrm{g}}^{(n)},\;
\bm{F}_{i,\mathrm{s}}^{(n)},\;
\bm{F}_{i,\mathrm{t}}^{(n)},\;
\bm{F}_{i,\mathrm{a}}^{(n)}
\big] \in \mathbb{R}^{13},
\]
where \( \bm{F}_{i,\mathrm{g}}^{(n)} \), \( \bm{F}_{i,\mathrm{s}}^{(n)} \), \( \bm{F}_{i,\mathrm{t}}^{(n)} \), and \( \bm{F}_{i,\mathrm{a}}^{(n)} \) denote the geometric, spatial, structural, and intensity features, respectively. These feature groups jointly capture the size and shape, spatial location, morphological structure, and radiative properties of each filament fragment, thereby enabling a more comprehensive and discriminative node representation. Stacking all node features yields the node feature matrix
\[
\bm{X}^{(n)} = \left[
\bm{F}_1^{(n)},\; \bm{F}_2^{(n)},\; \cdots,\; \bm{F}_{N_n}^{(n)}
\right]^{\top}
\in \mathbb{R}^{N_n \times 13},
\]
where \( N_n \) denotes the number of nodes in the \( n \)th graph. The constructed feature vectors serve as the initial node representations, i.e., \( \bm{h}_v^{(0)} = \bm{F}_v^{(n)} \), and are further refined through iterative message passing in the GNN.

\subsection{Edge Feature Representation} 
While node features can characterize the intrinsic properties of individual solar filament fragments, they are insufficient to describe the relationships between fragments. Therefore, we explicitly construct edge features to model pairwise interactions, thereby encoding both spatial proximity and structural similarity between connected nodes. For each edge \( (i,j)\in E^{(n)} \) (with undirected connectivity), we define an edge feature extraction function:
\[
\bm{e}_{ij}^{(n)} = \phi\big(\bm{F}_{i}^{(n)}, \bm{F}_{j}^{(n)}, \bm{c}_{i}^{(n)}, \bm{c}_{j}^{(n)}\big) \in \mathbb{R}^{q},
\]
where \( \bm{F}^{(n)} \) denote the node features, and \( \bm{c}^{(n)} = (c^x,\, c^y) \) represents the centroid of fragment \( P^{(n)} \). In this work, the mapping function $\phi(\cdot)$ is explicitly instantiated.
It is constructed using a set of geometric relations and structural descriptors derived from node features, as detailed below. The normalized distance is defined as
\[
d_{ij}^{(n)} = \frac{\left\| \bm{c}_i^{(n)} - \bm{c}_j^{(n)} \right\|_2}{D^{(n)}},
\]
where the scale factor \(D^{(n)}\)  is defined as
\[
D^{(n)} = \sqrt{H_{\mathrm{range}}^2 + W_{\mathrm{range}}^2},
\]
where \( H_{\mathrm{range}} \) and \( W_{\mathrm{range}} \) are computed from the spatial extent of all fragment centroids, rather than the full image size, to provide a data-adaptive normalization that better reflects the local distribution of fragments. 
The absolute centroid differences describe the coordinate-wise spatial separation between fragments:
\[
\Delta x_{ij} = \left| c_i^x - c_j^x \right|, \quad
\Delta y_{ij} = \left| c_i^y - c_j^y \right|,
\]
and the normalized coordinate differences are given by
\[
\widetilde{\Delta x}_{ij}^{(n)} = \frac{\Delta x_{ij}}{D^{(n)}}, \quad
\widetilde{\Delta y}_{ij}^{(n)} = \frac{\Delta y_{ij}}{D^{(n)}}.
\]
Structural similarity between fragments is quantified by the absolute difference between their node features:
\[
\Delta \bm{F}_{ij}^{(n)} = \left| \bm{F}_i^{(n)} - \bm{F}_j^{(n)} \right|.
\]
A reduced feature difference vector is retained from $\Delta \bm{F}_{ij}^{(n)}$, yielding
\[
\widetilde{\Delta \bm{F}}_{ij}^{(n)} =
\left[
\Delta A_{ij},\;
\Delta \mathrm{Peri}_{ij},\;
\Delta L_{ij},\;
\Delta \epsilon_{ij}
\right]
\]
where \( \Delta A_{ij} \), \( \Delta \mathrm{Peri}_{ij} \), \( \Delta L_{ij} \), and \( \Delta \epsilon_{ij} \) denote the absolute differences in area, perimeter, skeleton length, and eccentricity, respectively. By combining spatial and structural descriptors, the final edge feature is defined as
\[
\bm{e}_{ij}^{(n)} =
\left[
d_{ij}^{(n)},\;
\widetilde{\Delta x}_{ij}^{(n)},\;
\widetilde{\Delta y}_{ij}^{(n)},\;
\widetilde{\Delta \bm{F}}_{ij}^{(n)}
\right]
\in \mathbb{R}^{7}.
\]
Thus, each edge is represented by a 7-dimensional feature vector encoding both spatial proximity and structural consistency between fragments. Stacking all edge features yields
\[
\bm{E}^{(n)} \in \mathbb{R}^{|E^{(n)}| \times 7}.
\]
Since the graph is constructed as undirected, the edge representation should be invariant to node ordering. 
To this end, all components are defined using symmetric operations (e.g., absolute differences and Euclidean distance), which leads to
\[
\bm{e}_{ij}^{(n)} = \bm{e}_{ji}^{(n)}.
\]
This symmetric formulation ensures that edge features are consistent with the undirected graph structure.

\subsection{Solar Filament Fragment Association Prediction}

\begin{figure}[!htbp]
\centering
\plotone{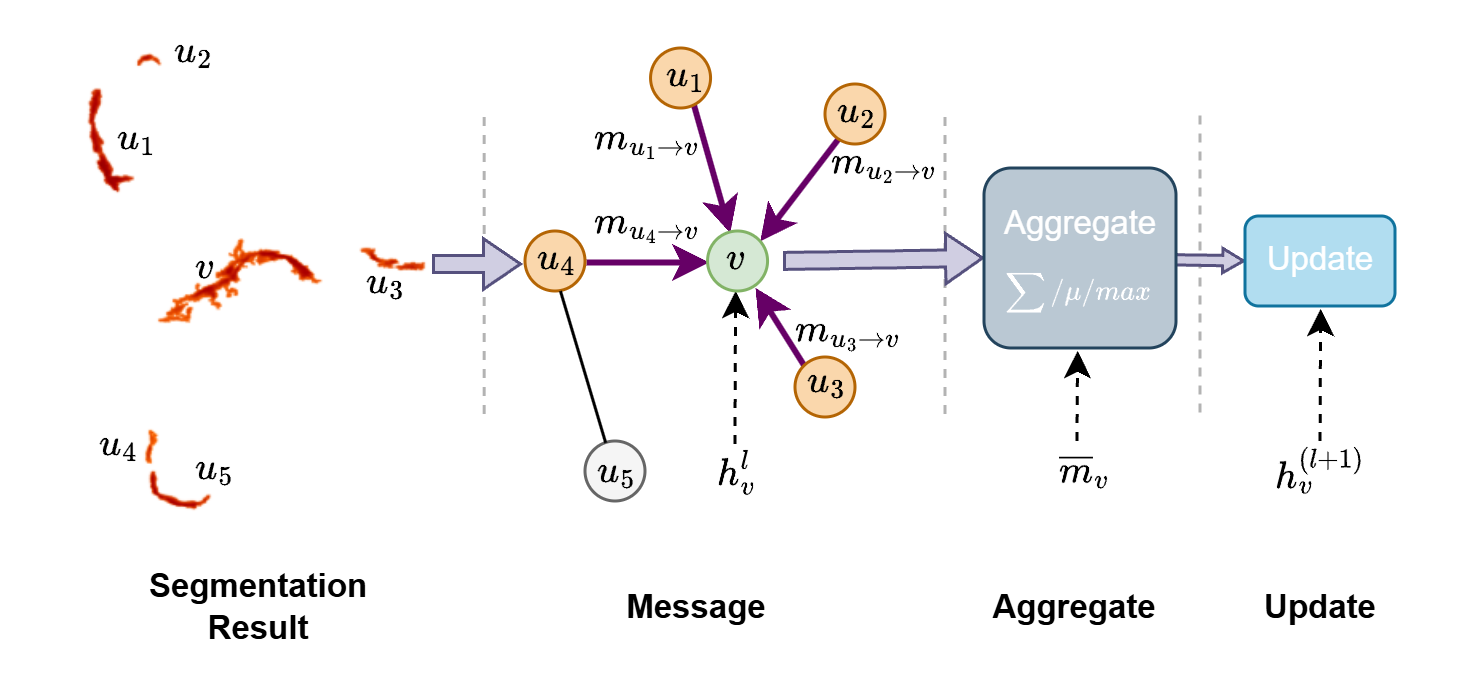}
\caption{Illustration of the message-passing mechanism in the graph neural network. Filament fragments are represented as nodes, where neighboring nodes transmit messages to a node. These messages are aggregated via an aggregation function, and the node features are subsequently updated to obtain refined embeddings.}
\label{fig:m_a_u}
\end{figure}

In solar H$\alpha$ images, filament fragments belonging to the same physical structure typically exhibit strong local consistency, including spatial proximity, similar morphology, and consistent orientation.  However, under complex or atypical conditions, individual fragments may appear ambiguous or incomplete, making it difficult to determine their relationships. Therefore, it is essential to incorporate contextual information from neighboring fragments when inferring their associations (Figure~\ref{fig:m_a_u}). To capture such contextual dependencies among fragments, we construct a graph representation over filament segments. In the constructed graph \( G^{(n)} = (V^{(n)}, E^{(n)}) \),  the filament fragment association problem is formulated as an edge-level prediction task aimed at recovering the underlying connectivity among fragments belonging to the same solar filament. Instead of directly performing global grouping, we adopt an edge-centric formulation, where for each edge \( (i,j) \in E^{(n)} \), the model predicts the probability that two candidate fragments belong to the same filament. This formulation enables the global filament structure to be reconstructed from local pairwise predictions over graph edges.

To model such dependencies, we adopt the message-passing neural network (MPNN) framework \citep{gilmer2017neural}, which provides a unified mechanism for propagating and aggregating information over graph-structured data. In this framework, each node corresponds to a solar filament fragment with an associated feature vector, while edges represent candidate relationships between fragments. Similar to convolutional neural networks (CNNs), MPNNs employ stacked layers to update feature representations progressively. However, unlike CNNs that operate on regular grid structures, the multilayer design in MPNNs enables iterative message passing over graph neighborhoods, allowing each fragment to incorporate contextual information from multihop neighbors. At each layer, node representations are iteratively updated by aggregating messages from their neighboring nodes (Figure~\ref{fig:m_a_u}). Specifically, for a node \( v \), its representation at layer \( l+1 \) is computed as
\[
\bm{h}_v^{(l+1)} = U^{(l)}\left( \bm{h}_v^{(l)}, \; \sum_{u \in \mathcal{N}(v)} M^{(l)}\left( \bm{h}_v^{(l)}, \bm{h}_u^{(l)}, \bm{e}_{uv} \right) \right),
\]
where \( \bm{h}_v^{(l)} \) denotes the feature representation of node \( v \) at layer \( l \), \( \bm{e}_{uv} \) denotes the feature associated with edge \( (u,v) \), \( M^{(l)}(\cdot) \) is a message function that computes messages from neighboring nodes based on their features and pairwise relations, the summation operator aggregates incoming messages from the neighborhood, and \( U^{(l)}(\cdot) \) is a update function that incorporates the aggregated messages to update the node representation. This formulation enables each fragment to incorporate contextual cues from its local neighborhood. Moreover, by stacking multiple message-passing layers, the receptive field of each node is progressively expanded, allowing the model to capture long-range dependencies and approximate global structural context, thereby improving the robustness of association prediction under ambiguous conditions. After \( L \) layers of message passing, each node \( v \) obtains a final representation \( \bm{h}_v^{(L)} \), which encodes both local and multihop contextual information from the graph. Based on these learned node embeddings, we define an edge-level prediction function \( \sigma (\cdot) \) to estimate the probability that two connected nodes belong to the same filament. Specifically, for an edge \( (u,v) \in E \),
\[
p_{uv}=\sigma \bigl( \bm{h}_{u}^{(L)},\bm{h}_{v}^{(L)},\bm{e}_{uv} \bigr) ,
\]
where \( \bm{e}_{uv} \) denotes the edge feature associated with the node pair. This design enhances node representations through multihop propagation, enabling association predictions to leverage not only local fragment information but also contextual cues propagated from other fragments across the structure. Compared to other methods, this approach provides a more comprehensive model of filament connectivity, resulting in predictions that better align with the physical continuity and overall structure of solar filaments.

While the MPNN framework provides a general formulation for modeling graph-structured data, its practical effectiveness depends on the specific design of the message, aggregation, and update functions. In practice, several GNN architectures can be interpreted as specific instantiations of the MPNN framework, including graph convolutional networks (GCN) \citep{kipf2017semi} and graph attention networks (GAT) \citep{velivckovic2018graph}. GCN adopts a normalized aggregation scheme, where node representations are updated through normalized aggregation of neighboring node features:
\[
\bm{h}_v^{(l+1)} = \sigma \left( \sum_{u \in \mathcal{N}(v) \cup \{v\}} \frac{1}{\sqrt{\widetilde{d}_u \widetilde{d}_v}} \bm{W}^{(l)} \bm{h}_u^{(l)} \right),
\]
where \( \widetilde{d}_v \) denotes the degree of node \( v \) after adding self-loops, and \( \bm{W}^{(l)} \) is a learnable weight matrix. This formulation leads to stable but relatively uniform feature propagation across neighbors, as it assigns fixed normalized weights and lacks the ability to adaptively differentiate the importance of different fragments. However, this uniform weighting scheme limits the model's capacity to capture the varying relevance of neighboring nodes under complex conditions. To address this limitation, GAT replaces the fixed aggregation weights with an attention mechanism that assigns adaptive importance to different neighbors:
\[
\bm{h}_v^{(l+1)} = \sigma \left( \sum_{u \in \mathcal{N}(v) \cup \{v\}} \alpha_{uv}^{(l)} \bm{W}^{(l)} \bm{h}_u^{(l)} \right),
\]
where the attention coefficients \( \alpha_{uv}^{(l)} \) are computed as
\[
\alpha_{uv}^{(l)} = \mathrm{softmax}_u \big( a^{(l)}( [\bm{h}_v^{(l)} \, \| \, \bm{h}_u^{(l)}] ) \big),
\]
and \( a^{(l)}(\cdot) \) is a learnable attention function. This graph-based formulation allows each fragment to use information from both its immediate neighbors and more distant fragments connected in the graph. It is well suited for solar filament fragment association, where neighboring fragments can be unreliable due to fragmentation, noise, and occlusion.

\begin{figure}[!htbp]
\centering
\includegraphics[width=1\textwidth]{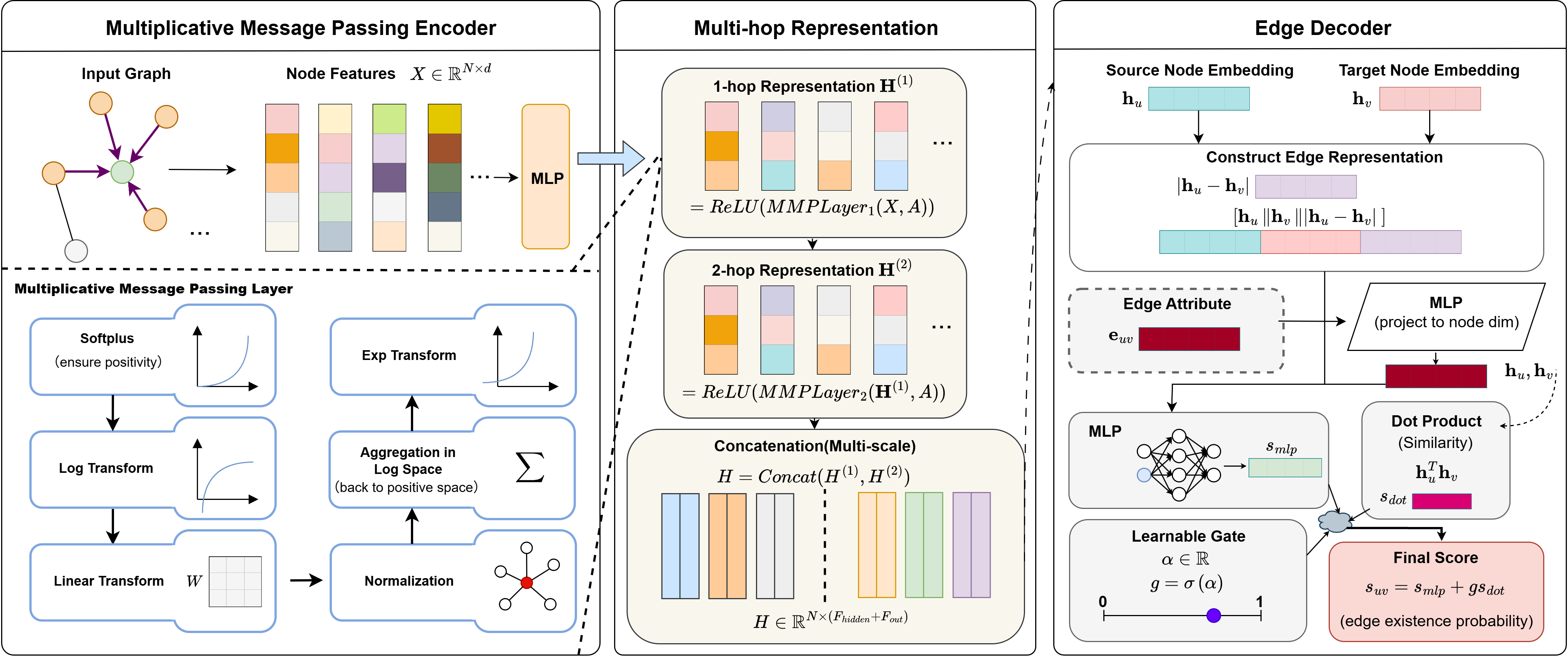}
\caption{Architecture of the proposed graph neural network. The encoder employs a multiplicative message-passing scheme to learn node embeddings. Multihop feature layers are combined to capture multiscale structural information. The edge decoder leverages node embeddings and edge attributes, together with a learnable gating mechanism, to predict edge existence probabilities.}
\label{fig:model}
\end{figure}

While GCN and GAT can effectively model contextual information over graph neighborhoods, their additive aggregation mechanism mainly combines information from neighboring nodes through summation. In this aggregation scheme, a highly consistent neighbor may compensate for inconsistent neighbors, making the representation less sensitive to the disagreement among multiple fragments. In the solar filament segment association task, true connectivity typically depends on multiple fragments simultaneously exhibiting consistent spatial and morphological characteristics, rather than being determined by a single strongly matched neighbor. As a result, purely additive aggregation may limit the model's ability to capture such joint consistency among neighboring fragments. This property is similar to the motivation of geometric averaging, where a low-valued feature can reduce the overall representation instead of being completely compensated by other features. To capture this property, we propose a multiplicative message-passing mechanism that replaces traditional additive aggregation with a weighted geometric aggregation scheme (Figure~\ref{fig:model}). For a node \(v\), the update rule is defined as follows:
\[
\bm{h}_v^{(l+1)} =
\exp \left(
\sum_{u \in \mathcal{N}(v) \cup \{v\}}
\frac{1}{\sqrt{\widetilde{d}_u \widetilde{d}_v}}
\bm{W}^{(l)}
\ln\big( \phi(\bm{h}_u^{(l)}) \big)
\right).
\]
Message passing is carried out in the logarithmic domain, followed by an exponential mapping back to the original space, resulting in a numerically stable implementation of multiplicative aggregation. This formulation is equivalent to a weighted geometric aggregation over neighboring features, which can be explicitly written in multiplicative form as
\[
\bm{h}_v^{(l+1)} =
\prod_{u \in \mathcal{N}(v) \cup \{v\}}
\left( \phi(\bm{h}_u^{(l)}) \right)^{\frac{1}{\sqrt{\tilde{d}_u \tilde{d}_v}}\bm{W}^{(l)}}.
\]

Compared with additive aggregation, the proposed weighted geometric aggregation reduces the compensation effect caused by inconsistent neighboring features. Therefore, when one or more neighboring fragments exhibit incompatible characteristics, the aggregated representation is decreased, while consistent neighboring fragments jointly contribute to a stronger representation. By integrating this multiplicative aggregation mechanism into the message-passing process, the proposed method provides a representation that better reflects the local consistency among associated solar filament fragments. This design is particularly beneficial in challenging scenarios where individual segments are incomplete or ambiguous, since reliable association requires agreement from multiple neighboring fragments rather than isolated evidence.

Based on the learned node representations (Figure~\ref{fig:model}), we construct a pairwise feature for each candidate edge $(u,v)$ by combining node embeddings and their relational difference, and directly map it to a compatibility score:
\[
s_{uv} =
f_{\theta} \!\left(
\mathrm{concat}\!\left(
\bm{h}_u^{(l)},\;
\bm{h}_v^{(l)},\;
\left| \bm{h}_u^{(l)} - \bm{h}_v^{(l)} \right|,\;
\psi(\bm{e}_{uv})
\right)
\right)
\;+\;
\gamma \, (\bm{h}_u^{(l)})^{\top}\bm{h}_v^{(l)},
\]
where $f_{\theta}(\cdot)$ denotes a multilayer perceptron and $\gamma$ is a learnable gating parameter. The score is optimized using a binary cross-entropy loss with logits. The overall design encourages the model to assign high scores to edges only when multiple adjacent segments exhibit consistent characteristics, thereby improving the accuracy of filament fragment association.

\section{Experiments} 
\label{sec:exp}

\subsection{Data Description}

Since there is currently no publicly available dataset specifically designed for filament fragment association, we construct a dedicated dataset for this study. The original H\(\alpha\) images are obtained from observations at the Kanzelhöhe Solar Observatory (KSO). The dataset contains approximately 100 H\(\alpha\) images, covering various filament structures under different solar conditions. From these images, we extract approximately 1700 filament fragments. Filament regions are first obtained using a deep learning--based segmentation method \citep{zhu2025flat}, which is widely adopted in recent solar image analysis studies. Since automatic segmentation cannot always achieve perfect accuracy, the generated masks are further visually inspected and manually corrected using the LabelMe tool \citep{russell2008labelme}. This manual refinement step aims to remove false detections, correct missed filament regions, and improve the accuracy of filament boundaries. During this annotation process, fragments originating from the same physical filament are assigned the same group identifier, which is used as the ground truth for association prediction. To improve numerical stability, logarithmic scaling and global normalization are applied to selected features. Candidate edges are constructed using a spatial $k$-nearest neighbor (kNN) strategy with a distance constraint to remove long-range connections and reduce computational cost. For each candidate edge, pairwise features are computed, including normalized spatial distance, relative position \((dx, dy)\), and differences in key fragment attributes such as area, perimeter, skeleton length, and eccentricity. Edge labels are defined as binary labels: an edge is positive if the two connected fragments have the same group identifier (i.e., belong to the same filament), and negative otherwise. All node features are standardized on a per-image basis to have zero mean and unit variance. Each image is then transformed into an attributed graph, forming a dataset of graph samples for training and evaluating the proposed model.

\subsection{Experimental Setup}

All experiments were implemented using PyTorch and PyTorch Geometric. Model training was conducted on a GPU when available, although CPU execution is also feasible. The constructed dataset was randomly shuffled and partitioned at the graph level, with 80\% of the samples used for training, and 20\% for validation and testing. A GNN framework is used for solar filament fragment association prediction. The input dimension is determined by the node features (13 dimensions). The model uses a 64-dimensional hidden layer and a 128-dimensional output embedding. Edge features constructed during graph generation are incorporated into edge prediction. Given an input graph, the model outputs the probability of candidate edges between filament fragments. The model is trained for 100 epochs using the Adam optimizer with a learning rate of \(1 \times 10^{-3}\). The model is optimized using the binary cross-entropy loss.

\subsection{Results}
\subsubsection{Overall Performance Comparison}

Solar filament fragment association is not a simple independent prediction task. The edge labels are related to each other and should form a consistent connection structure. If two fragments are connected, this often affects other nearby connections. Fragments from the same filament are expected to form a continuous structure. Because of this, the correctness of one edge does not only depend on the two fragments themselves. It also depends on how well this edge agrees with nearby connections. This makes the task different from standard independent classification, and requires the model to consider consistency in the local structure. The initial graph construction is therefore important, as it decides which fragments can interact. In this work, candidate edges are built using a kNN method. Each fragment is connected to its $k$-nearest neighbors in space. The value of $k$ controls the graph density. A small $k$ keeps only local connections. A large $k$ adds more edges, but may also introduce noisy or unclear relations. To study this effect, we report results under different settings, with $k=5$, 10, and 15. All results are obtained using 5-fold cross-validation, and the mean and standard deviation are reported.

Table~\ref{tab:tb_result} reports the performance of different models under different kNN graph constructions, evaluated using the area under the ROC curve (AUC) and average precision (AP). Our method achieves the best results in all settings, as indicated by the bold values. For $k=5$, it reaches an AUC of 0.854 and an AP of 0.800. GAT and GCN obtain 0.838/0.783 and 0.835/0.773, respectively. The gain in AP is important because AP reflects performance under class imbalance. Compared with GCN, the method improves AP by about 2.7 percentage points. This shows better ranking of positive edges. When $k$ increases to 10, the gap becomes larger. It achieves the highest AUC of 0.867, while GAT and GCN reach 0.842 and 0.832. The improvement over GCN is about 3.5 percentage points. This shows that it makes better use of additional neighbors. In comparison, additive models are more affected by less relevant neighbors. For $k=15$, all methods show a slight drop or become stable. The proposed method still performs the best. The AP is 0.785, compared with 0.762 for GCN and 0.747 for GAT. A larger $k$ introduces more ambiguous connections. It is less affected by this noise. The multiplicative aggregation reduces the impact of inconsistent neighbors and keeps the performance stable.

\begin{table}[!htbp] \small
\centering
\caption{Performance under Different Initial kNN Graph Constructions}
\label{tab:tb_result}
\footnotesize
\begin{tabular}{lcccccc}
\hline\hline
Models 
& AUC ($k=5$) 
& AP ($k=5$) 
& AUC ($k=10$) 
& AP ($k=10$) 
& AUC ($k=15$) 
& AP ($k=15$)  \\
\hline
MLP & 0.808 $\pm$ 0.013 & 0.732 $\pm$ 0.038 & 0.817 $\pm$ 0.012 & 0.722 $\pm$ 0.053 & 0.811 $\pm$ 0.015 & 0.712 $\pm$ 0.045 \\
GCN & 0.835 $\pm$ 0.022 & 0.773 $\pm$ 0.060 & 0.832 $\pm$ 0.017 & 0.747 $\pm$ 0.060 & 0.835 $\pm$ 0.022 & 0.762 $\pm$ 0.057 \\
GAT & 0.838 $\pm$ 0.014 & 0.783 $\pm$ 0.057 & 0.842 $\pm$ 0.027 & 0.766 $\pm$ 0.071 & 0.830 $\pm$ 0.035 & 0.747 $\pm$ 0.077 \\
SAGE & 0.810 $\pm$ 0.025 & 0.741 $\pm$ 0.058 & 0.799 $\pm$ 0.037 & 0.699 $\pm$ 0.068 & 0.801 $\pm$ 0.021 & 0.699 $\pm$ 0.069 \\
\textbf{OURS} & \textbf{0.854 $\pm$ 0.010} & \textbf{0.800 $\pm$ 0.036} & \textbf{0.867 $\pm$ 0.014} & \textbf{0.792 $\pm$ 0.054} & \textbf{0.859 $\pm$ 0.011} & \textbf{0.785 $\pm$ 0.048} \\
\hline
\end{tabular}
\end{table}

The variation across different values of $k$ shows a clear trend. For the multilayer perceptron (MLP), performance does not consistently improve as $k$ increases. This is expected, since MLP does not use graph structure. The changes mainly come from different edge samples, not from structural information. For GCN and GAT, a larger $k$ does not always help. Both models show a drop at $k=15$. Additive aggregation combines information from all neighbors. When the graph becomes dense, less relevant connections are also included. This weakens useful signals. In contrast, the proposed method benefits from a moderate increase in $k$ (from 5 to 10). The performance drops less at $k=15$. This shows that the multiplicative aggregation works differently. It focuses more on consistency across neighbors, instead of simply summing their features. Inconsistent neighbors have less influence, which reduces their negative effect.

Another notable result is the variance across folds. Our model consistently shows lower standard deviation compared to other models. For example, at $k=5$, the AUC standard deviation of the proposed method is 0.010,  while GCN reaches 0.022 and GraphSAGE (Graph Sample and Aggregate, \citeauthor{hamilton2017inductive} \citeyear{hamilton2017inductive}) reaches 0.025. This shows more stable performance under different data splits. This stability is important in this task because the dataset is relatively small. A model that is sensitive to data partitioning may not generalize well. The lower variance suggests that the learned representation captures more stable structural patterns rather than overfitting to specific samples.

\begin{figure}[!htbp]
\centering
\epsscale{0.72}
\plotone{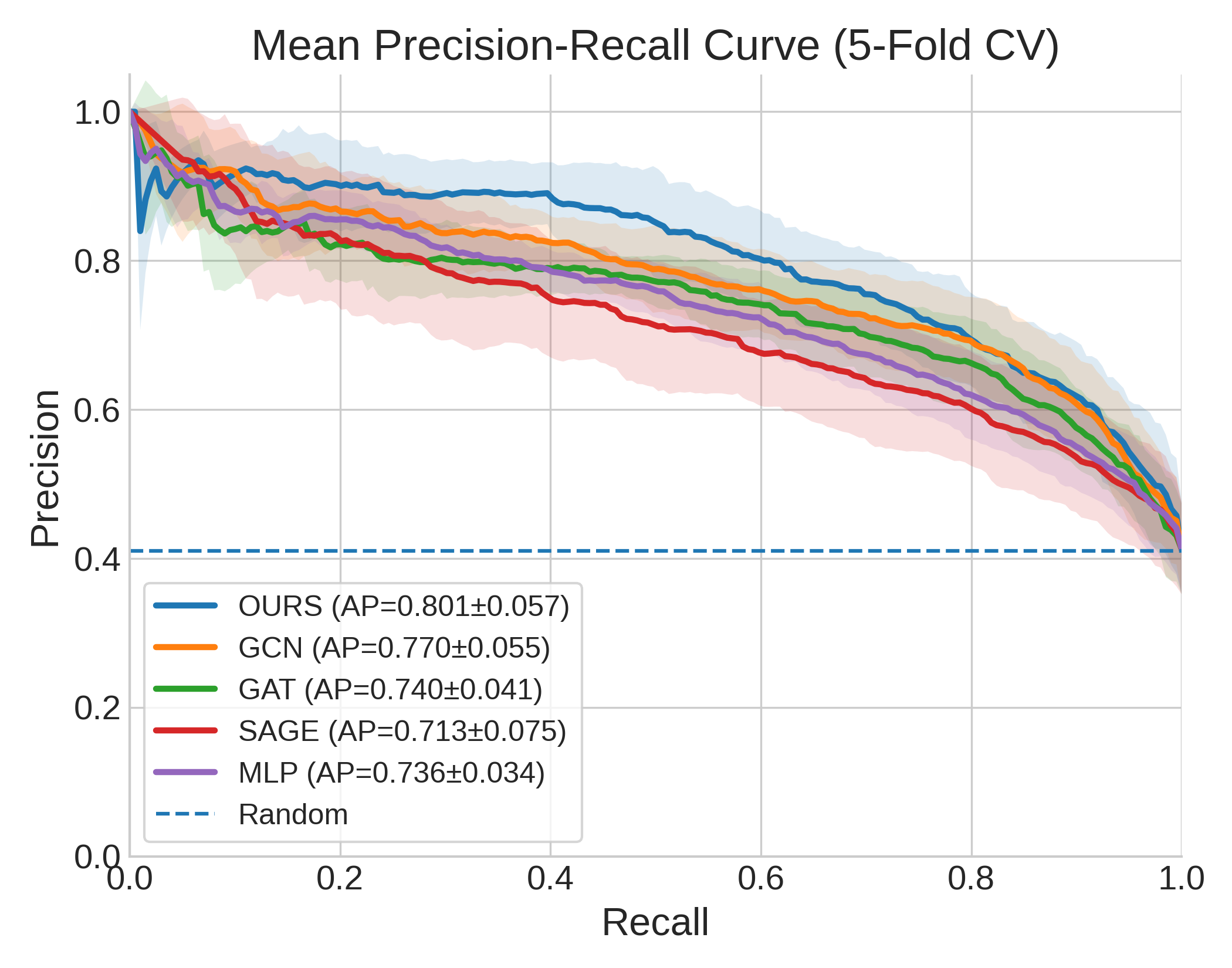}
\caption{Mean precision–recall curves with standard deviation bands over 5-fold cross-validation. The proposed method achieves the highest average precision (AP) among all compared models (GCN, GAT, SAGE, and MLP), indicating improved performance in filament fragment association.}
\label{fig:mean_pr_curve}
\end{figure}

\begin{figure}[!htbp]
\centering
\includegraphics[width=1\textwidth]{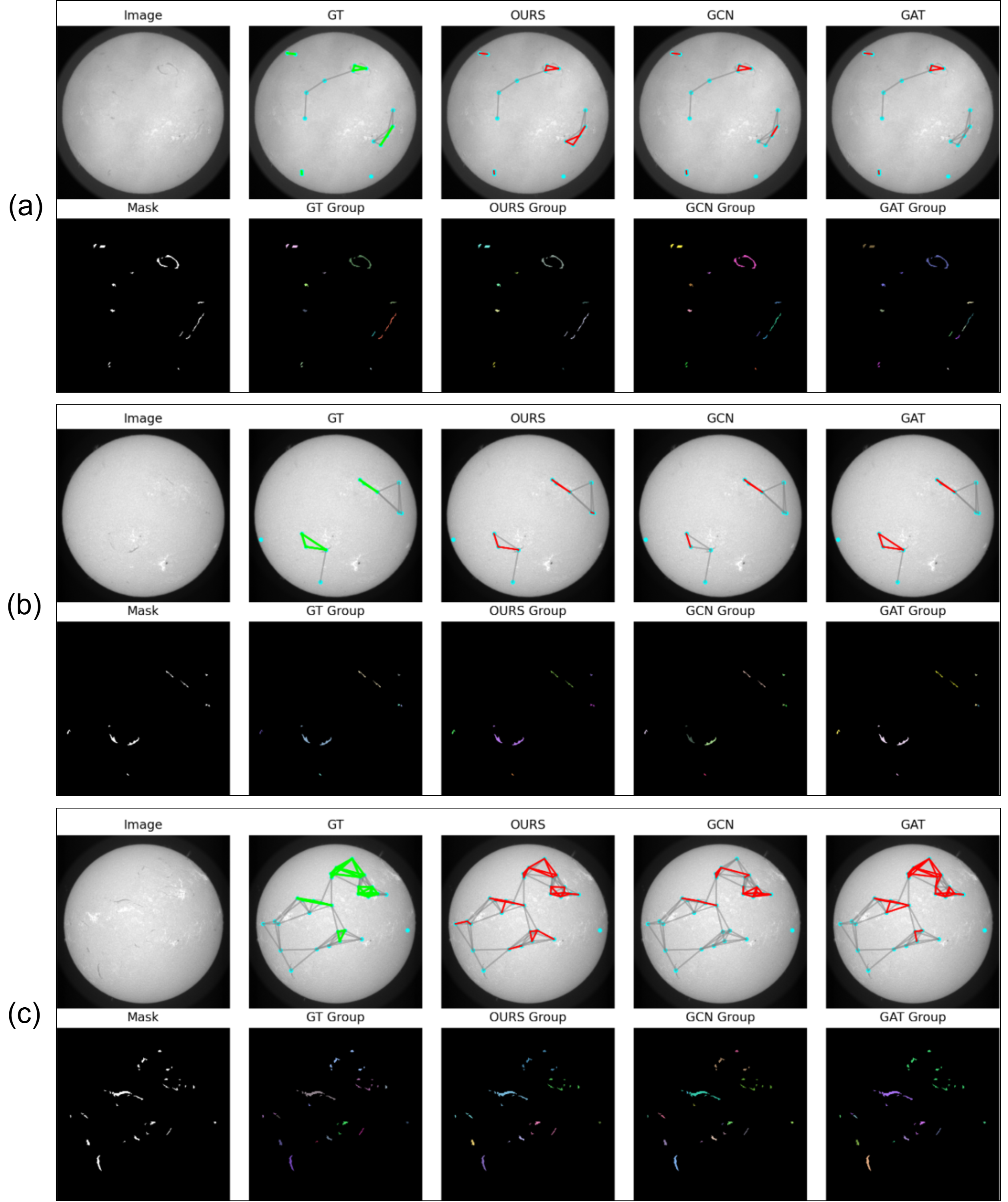}
\caption{Comparison of filament fragment association results. From left to right: original H$\alpha$ image, ground truth (GT), and predictions from different methods (OURS, GCN, GAT). The second row in each panel shows the corresponding segmentation masks and grouped fragment associations. The proposed method produces results that are closer to the ground truth.}
\label{fig:cmp_ours_gcn_gat}
\end{figure}

Figure~\ref{fig:mean_pr_curve} provides a detailed view of model performance across different recall levels. The curve of the method lies above all baselines over most of the recall range. The difference is clear in the mid-recall region (0.2--0.8), where practical decision thresholds usually lie. In this range, it achieves higher precision than GCN and GAT. At high recall (close to 1.0), all methods show lower precision, as more uncertain edges are included. The method still performs slightly better, which reflects a better ranking of edge scores. The shaded area (standard deviation) is also narrower, showing more consistent results across folds. The dashed horizontal line represents the random baseline. All methods perform better than this baseline, which shows that the task is learnable. The gap between the method and the baseline is the largest, indicating stronger discrimination ability. Figure~\ref{fig:cmp_ours_gcn_gat} shows visual comparisons. In the first row of each panel, the predicted connections are closer to the ground truth. The method correctly links fragments that form continuous filament structures, even when they are partially disconnected or spatially separated. In contrast, GCN and GAT miss some true connections and introduce incorrect ones. The second row in each panel highlights the grouping results. The method produces more coherent groups that better match the ground truth. The predicted groups preserve the continuity of filament structures. GCN and GAT produce more fragmented groups or merge unrelated fragments.

\subsubsection{Temporal Generalization Analysis}

Although random cross-validation is widely used, samples from close observation periods may share similar solar conditions. To further test the generalization ability of the models, we performed an additional temporal split experiment. All samples were sorted by their observation time, and five folds were constructed sequentially without random shuffling. In this setting, models were trained on earlier observations and tested on later observations, providing a stricter evaluation of temporal generalization.

Table~\ref{tab:temporal_split} shows the results of the temporal split experiment. Compared with the random cross-validation results in Table~\ref{tab:tb_result}, all methods show performance decreases. This is expected because temporal splitting may introduce larger variations in solar morphology and observation conditions between the training and testing samples. For the proposed method, the AUC decreases to 0.819 and the AP decreases to 0.751 (shown in bold). This result shows that filament fragment association becomes more difficult when the model needs to predict samples from different observation periods.

\begin{table}[!htbp]
\centering
\caption{Performance Comparison under Temporal Split Evaluation}
\label{tab:temporal_split}
\setlength{\tabcolsep}{12pt}
\begin{tabular}{lcc}
\hline
\hline
Models & AUC & AP  \\
\hline
MLP & $0.807\pm 0.045$ & $0.730\pm 0.075$  \\
GCN & $0.801\pm 0.039$ & $0.738\pm 0.061$  \\
GAT & $0.805\pm 0.044$ & $0.751\pm 0.077$  \\
SAGE & $0.716\pm 0.068$ & $0.621\pm 0.097$  \\
\textbf{OURS} & \textbf{0.819 $\pm$ 0.028} & \textbf{0.751 $\pm$ 0.071}\\
\hline
\end{tabular}
\end{table}

Despite this challenge, the proposed method still provides reliable performance under this stricter evaluation setting, achieving an AUC of 0.819 and an AP of 0.751. Compared with GCN and GAT, it improves the AUC by 0.018 and 0.014, respectively. The improvement is similar to that observed in the random split setting, suggesting that the advantage of the proposed method is not only due to temporal similarity between training and testing samples. Instead, it comes from the ability of the proposed approach to capture reliable fragment relationships. The performance variation, especially for AP, becomes larger under the temporal split setting. This is mainly caused by the changes in solar activity and observation conditions over different time periods. Nevertheless, the proposed method maintains competitive performance across different temporal folds, demonstrating its potential to generalize across different observation periods.

\subsubsection{Feature Contribution Analysis}

To investigate how the proposed model leverages different attributes of solar filament segments, we conducted feature permutation experiments. Unlike conventional feature ablation methods that alter the input feature space, feature permutation preserves the original feature space and model architecture while randomly disrupting the correspondence between feature values and segments. Since other feature groups remain unchanged, performance degradation after permutation more directly reflects the contribution of the permuted feature group. As shown in Table~\ref{tab:joint_permutation}, among all feature groups, spatial attributes contribute the most to the prediction. After permuting the position-related node and edge attributes, the AUC decreases to 0.719 and the AP decreases to 0.638. This large performance drop shows that spatial information provides the dominant cue for the model. This is also consistent with the fact that connected filament fragments usually have spatial continuity. Permuting size- and structure-related attributes also reduces the performance, but the decrease is smaller. The AUC values decrease to 0.826 and 0.830, respectively. This indicates that the size and structural properties of filament segments provide additional information for association prediction. The intensity-related attributes have the smallest effect, with the AUC remaining at 0.838 after permutation. This suggests that intensity information is helpful but is not the main factor. Overall, the permutation results show that the proposed model uses both spatial relationships and fragment-level properties. Spatial information provides the main signal, while size, structure, and intensity features provide complementary information for reliable filament fragment association.

\begin{table}[!htbp]
\centering
\caption{Joint Permutation Analysis of Different Feature Groups}
\label{tab:joint_permutation}
\begin{tabular}{lllcc}
\hline\hline
Feature group & Node attributes & Edge attributes & AUC & AP  \\
\hline
Position & $\tilde{c}_i^x,\tilde{c}_i^y$ & $d_{ij}^{(n)},\widetilde{\Delta x}_{ij}^{(n)},\widetilde{\Delta y}_{ij}^{(n)}$ & $0.7189\pm 0.0223$ & $0.6381\pm 0.0902$  \\
Size & $\log A_i,\log \mathrm{Peri}_i,\log w_i,\log h_i,r_i$ & $\Delta A_{ij},\Delta\mathrm{Peri}_{ij}$ & $0.8260\pm 0.0243$ & $0.7515\pm0.0692$  \\
Structure & $\log L_i,\epsilon_i,\sin\theta_i,\cos\theta_i$ & $\Delta L_{ij},\Delta\epsilon_{ij}$ & $0.8297\pm 0.0198$ & $0.7589\pm 0.0504$  \\
Intensity & $\hat{\mu}_i,\hat{\sigma}_i$ & None & $0.8377\pm 0.0189$ & $0.7588\pm 0.0582$  \\
\hline
\end{tabular}
\raggedright
\tablecomments{The corresponding node attributes and edge attributes are simultaneously permuted while preserving the graph topology.}
\end{table}

\subsubsection{Summary of Results}

The experimental results demonstrate the effectiveness of the proposed graph-based model for solar filament fragment association. The model achieves consistent improvements under different graph settings and evaluation schemes, showing that the learned fragment representations are robust and not strongly dependent on a specific data split. The temporal split experiment further shows that the model can generalize to later observation periods. In addition, the feature permutation analysis shows that spatial information provides the main information, while other fragment attributes provide complementary cues. These results indicate that combining spatial relationships with multiple fragment attributes is important for reliable solar filament fragment association.

\section{Discussion And Conclusion} 
\label{sec:dis_con}

Solar filaments often break into multiple fragments as they evolve, which makes automatic tracking difficult. In H$\alpha$ images, these fragments can appear separated in space because of projection effects and the 3D structure of the magnetic field, even though they belong to the same filament structure. In this work, we study the problem of filament fragment association using graph learning. Instead of using rule-based merging methods or instance segmentation, we treat each fragment as a node in a graph and infer whether two nodes should be connected as an edge prediction task. This strategy separates the feature extraction of each fragment from the modeling of its relationships. It allows the model to use both the properties of each fragment and the connections between fragments. This is useful when a single physical filament is split into several disconnected parts in the image.

A key observation from this study is that filament fragment association is fundamentally a structural consistency problem rather than a purely pairwise similarity problem. In real H$\alpha$ observations, fragments belonging to the same filament may exhibit significant spatial gaps, while visually similar fragments may belong to different filaments. This makes purely distance-based or appearance-based criteria unreliable. By introducing graph learning, the model is able to leverage multihop contextual information, which provides additional constraints beyond local pairwise cues. The proposed multiplicative aggregation mechanism further strengthens this modeling paradigm. Unlike additive message passing, which accumulates information linearly from neighbors, the multiplicative formulation imposes a consistency-driven interaction among neighboring nodes. In practice, this leads to a filtering effect: contributions from inconsistent neighbors are naturally suppressed when their representations deviate in feature space. The experimental results under different kNN settings support this interpretation. When the graph becomes denser (larger $k$), additive models such as GCN and GAT are more sensitive to noisy or weakly related edges, resulting in performance degradation. In contrast, the proposed method maintains more stable performance, indicating improved robustness to graph construction variations. Additional temporal split experiments further show that the performance advantage is maintained when training and testing samples come from different observation periods. This suggests that the effectiveness of the proposed method does not mainly result from temporal similarity between samples. Feature permutation analysis also shows that spatial relationships provide the most important information, while fragment size, structural properties, and intensity characteristics provide complementary cues.

Despite these advantages, several limitations remain. First, the current graph construction relies on a $k$-nearest neighbor strategy, which is purely geometric and does not incorporate physical constraints from solar magnetic fields. Second, although the multiplicative aggregation improves robustness, it may also reduce sensitivity to subtle but informative variations between fragments. Designing adaptive mechanisms that balance consistency and discriminability is a potential direction for improvement. Third, the current node and edge features are manually designed based on prior knowledge of fragment properties. Although these features provide meaningful physical descriptions, automatically learning more expressive representations from the data may further improve the model capability and reduce dependence on handcrafted feature engineering.

In summary, we propose a graph-based learning approach for the solar filament fragment association task. We treat each fragment as a node and build a graph to describe its connections. We design a GNN that updates node features through message passing with multiplicative aggregation. The model learns both the features of individual fragments and their relationships in the graph. Experimental results show that our method performs better than classical graph models such as GCN and GAT under different graph construction settings. It also produces more stable results across different data splits. These results show that graph-based learning is effective for modeling relationships between filament fragments, and that using multiplicative aggregation helps improve performance in this task.

\section*{Acknowledgments}
Our research is supported by the Talent Introduction Program of Xihua University (Z242123). It is also funded by National Key R\&D Program of China (2022YFF0503800), National Natural Science Foundation of China (12173049), and projects from National Astronomical Observatories, CAS (H242522, H242079).

%% pdflatex zgf26.tex
%% bibtex zgf26
%% pdflatex zgf26.tex
%% pdflatex zgf26.tex

\bibliography{zgf26}{}

\begin{thebibliography}{}
\expandafter\ifx\csname natexlab\endcsname\relax\def\natexlab#1{#1}\fi
\providecommand{\url}[1]{\href{#1}{#1}}
\providecommand{\dodoi}[1]{doi:~\href{http://doi.org/#1}{\nolinkurl{#1}}}
\providecommand{\doeprint}[1]{\href{http://ascl.net/#1}{\nolinkurl{http://ascl.net/#1}}}
\providecommand{\doarXiv}[1]{\href{https://arxiv.org/abs/#1}{\nolinkurl{https://arxiv.org/abs/#1}}}

\bibitem[{{Bernasconi} {et~al.}(2005){Bernasconi}, {Rust}, \&
  {Hakim}}]{bernasconi2005advanced}
{Bernasconi}, P.~N., {Rust}, D.~M., \& {Hakim}, D. 2005, \solphys, 228, 97,
  \dodoi{10.1007/s11207-005-2766-y}

\bibitem[{{Chen}(2008)}]{chen2008initiation}
{Chen}, P.~F. 2008, Journal of Astrophysics and Astronomy, 29, 179,
  \dodoi{10.1007/s12036-008-0023-0}

\bibitem[{{Chen} {et~al.}(2020){Chen}, {Xu}, \& {Ding}}]{chen2020some}
{Chen}, P.-F., {Xu}, A.-A., \& {Ding}, M.-D. 2020, Research in Astronomy and
  Astrophysics, 20, 166, \dodoi{10.1088/1674-4527/20/10/166}

\bibitem[{{Chen} {et~al.}(2023){Chen}, {Cheng}, {Chen}, {Dai}, \&
  {Ding}}]{chen2023observations}
{Chen}, Y., {Cheng}, X., {Chen}, J., {Dai}, Y., \& {Ding}, M. 2023, \apj, 959,
  67, \dodoi{10.3847/1538-4357/ad09d8}

\bibitem[{{Diercke} {et~al.}(2024){Diercke}, {Jarolim}, {Kuckein},
  {Gonz{\'a}lez Manrique}, {Ziener}, {Veronig}, {Denker}, {P{\"o}tzi},
  {Podladchikova}, \& {Pevtsov}}]{diercke2024universal}
{Diercke}, A., {Jarolim}, R., {Kuckein}, C., {et~al.} 2024, \aap, 686, A213,
  \dodoi{10.1051/0004-6361/202348314}

\bibitem[{{Fuller} {et~al.}(2005){Fuller}, {Aboudarham}, \&
  {Bentley}}]{fuller2005filament}
{Fuller}, N., {Aboudarham}, J., \& {Bentley}, R.~D. 2005, \solphys, 227, 61,
  \dodoi{10.1007/s11207-005-8364-1}

\bibitem[{{Gao} {et~al.}(2002){Gao}, {Wang}, \& {Zhou}}]{gao2002development}
{Gao}, J., {Wang}, H., \& {Zhou}, M. 2002, \solphys, 205, 93,
  \dodoi{10.1023/A:1013851808367}

\bibitem[{{Gilbert} {et~al.}(2000){Gilbert}, {Holzer}, {Burkepile}, \&
  {Hundhausen}}]{gilbert2000active}
{Gilbert}, H.~R., {Holzer}, T.~E., {Burkepile}, J.~T., \& {Hundhausen}, A.~J.
  2000, \apj, 537, 503, \dodoi{10.1086/309030}

\bibitem[{Gilmer {et~al.}(2017)Gilmer, Schoenholz, Riley, Vinyals, \&
  Dahl}]{gilmer2017neural}
Gilmer, J., Schoenholz, S.~S., Riley, P.~F., Vinyals, O., \& Dahl, G.~E. 2017,
  in International conference on machine learning, Pmlr, 1263--1272

\bibitem[{{Gopalswamy} {et~al.}(2003){Gopalswamy}, {Shimojo}, {Lu}, {Yashiro},
  {Shibasaki}, \& {Howard}}]{gopalswamy2003prominence}
{Gopalswamy}, N., {Shimojo}, M., {Lu}, W., {et~al.} 2003, \apj, 586, 562,
  \dodoi{10.1086/367614}

\bibitem[{{Guo} {et~al.}(2022){Guo}, {Yang}, {Feng}, {Bai}, {Liang}, \&
  {Dai}}]{guo2022solar}
{Guo}, X., {Yang}, Y., {Feng}, S., {et~al.} 2022, \solphys, 297, 104,
  \dodoi{10.1007/s11207-022-02019-z}

\bibitem[{Hamilton {et~al.}(2017)Hamilton, Ying, \&
  Leskovec}]{hamilton2017inductive}
Hamilton, W., Ying, Z., \& Leskovec, J. 2017, Advances in neural information
  processing systems, 30

\bibitem[{{Hao} {et~al.}(2015){Hao}, {Fang}, {Cao}, \&
  {Chen}}]{hao2015statistical}
{Hao}, Q., {Fang}, C., {Cao}, W., \& {Chen}, P.~F. 2015, \apjs, 221, 33,
  \dodoi{10.1088/0067-0049/221/2/33}

\bibitem[{{Hao} {et~al.}(2013){Hao}, {Fang}, \& {Chen}}]{hao2013developing}
{Hao}, Q., {Fang}, C., \& {Chen}, P.~F. 2013, \solphys, 286, 385,
  \dodoi{10.1007/s11207-013-0285-9}

\bibitem[{{Hou} {et~al.}(2023){Hou}, {Li}, {Li}, {Su}, {Qiu}, {Yang}, {Yang},
  {Li}, {Guo}, {Hou}, {Song}, {Bai}, {Zhou}, {Ding}, {Gan}, \&
  {Deng}}]{hou2023partial}
{Hou}, Y., {Li}, C., {Li}, T., {et~al.} 2023, \apj, 959, 69,
  \dodoi{10.3847/1538-4357/ad08bd}

\bibitem[{{Jiang} {et~al.}(2014){Jiang}, {Yang}, {Wang}, {Ji}, {Liu}, {Li}, \&
  {Li}}]{jiang2014interaction}
{Jiang}, Y., {Yang}, J., {Wang}, H., {et~al.} 2014, \apj, 793, 14,
  \dodoi{10.1088/0004-637X/793/1/14}

\bibitem[{{Jing} {et~al.}(2004){Jing}, {Yurchyshyn}, {Yang}, {Xu}, \&
  {Wang}}]{jing2004relation}
{Jing}, J., {Yurchyshyn}, V.~B., {Yang}, G., {Xu}, Y., \& {Wang}, H. 2004,
  \apj, 614, 1054, \dodoi{10.1086/423781}

\bibitem[{{Joshi} {et~al.}(2010){Joshi}, {Srivastava}, \&
  {Mathew}}]{joshi2010automated}
{Joshi}, A.~D., {Srivastava}, N., \& {Mathew}, S.~K. 2010, \solphys, 262, 425,
  \dodoi{10.1007/s11207-010-9528-1}

\bibitem[{{Joshi} {et~al.}(2014){Joshi}, {Magara}, \&
  {Inoue}}]{joshi2014formation}
{Joshi}, N.~C., {Magara}, T., \& {Inoue}, S. 2014, \apj, 795, 4,
  \dodoi{10.1088/0004-637X/795/1/4}

\bibitem[{{Joshi} {et~al.}(2023){Joshi}, {Luna}, {Schmieder},
  {Moreno-Insertis}, \& {Chandra}}]{joshi2023interaction}
{Joshi}, R., {Luna}, M., {Schmieder}, B., {Moreno-Insertis}, F., \& {Chandra},
  R. 2023, \aap, 672, A15, \dodoi{10.1051/0004-6361/202245647}

\bibitem[{Kipf \& Welling(2017)}]{kipf2017semi}
Kipf, T.~N., \& Welling, M. 2017, in International Conference on Learning
  Representations

\bibitem[{{Labrosse} {et~al.}(2010){Labrosse}, {Dalla}, \&
  {Marshall}}]{labrosse2010automatic}
{Labrosse}, N., {Dalla}, S., \& {Marshall}, S. 2010, \solphys, 262, 449,
  \dodoi{10.1007/s11207-009-9492-9}

\bibitem[{{Liu} {et~al.}(2021){Liu}, {Song}, {Lin}, \& {Wang}}]{liu2021solar}
{Liu}, D., {Song}, W., {Lin}, G., \& {Wang}, H. 2021, \solphys, 296, 176,
  \dodoi{10.1007/s11207-021-01920-3}

\bibitem[{{Liu} {et~al.}(2024){Liu}, {Ruan}, {Schmieder}, {Guo}, {Chen},
  {Zheng}, {Su}, \& {Wang}}]{liu2024filament}
{Liu}, Y., {Ruan}, G.~P., {Schmieder}, B., {et~al.} 2024, \aap, 687, A130,
  \dodoi{10.1051/0004-6361/202449774}

\bibitem[{{Mackovjak} {et~al.}(2021){Mackovjak}, {Harman},
  {Maslej-Kre{\v{s}}{\v{n}}{\'a}kov{\'a}}, \& {Butka}}]{mackovjak2021scss}
{Mackovjak}, {\v{S}}., {Harman}, M., {Maslej-Kre{\v{s}}{\v{n}}{\'a}kov{\'a}},
  V., \& {Butka}, P. 2021, \mnras, 508, 3111, \dodoi{10.1093/mnras/stab2536}

\bibitem[{Russell {et~al.}(2008)Russell, Torralba, Murphy, \&
  Freeman}]{russell2008labelme}
Russell, B.~C., Torralba, A., Murphy, K.~P., \& Freeman, W.~T. 2008,
  International journal of computer vision, 77, 157

\bibitem[{{Schmieder} {et~al.}(2013){Schmieder}, {D{\'e}moulin}, \&
  {Aulanier}}]{schmieder2013solar}
{Schmieder}, B., {D{\'e}moulin}, P., \& {Aulanier}, G. 2013, Advances in Space
  Research, 51, 1967, \dodoi{10.1016/j.asr.2012.12.026}

\bibitem[{{Shih} \& {Kowalski}(2003)}]{shih2003automatic}
{Shih}, F.~Y., \& {Kowalski}, A.~J. 2003, \solphys, 218, 99,
  \dodoi{10.1023/B:SOLA.0000013052.34180.58}

\bibitem[{{Sinha} {et~al.}(2019){Sinha}, {Srivastava}, \&
  {Nandy}}]{sinha2019solar}
{Sinha}, S., {Srivastava}, N., \& {Nandy}, D. 2019, \apj, 880, 84,
  \dodoi{10.3847/1538-4357/ab2239}

\bibitem[{Veli{\v{c}}kovi{\'c} {et~al.}(2018)Veli{\v{c}}kovi{\'c}, Cucurull,
  Casanova, Romero, Li{\`o}, \& Bengio}]{velivckovic2018graph}
Veli{\v{c}}kovi{\'c}, P., Cucurull, G., Casanova, A., {et~al.} 2018, in
  International Conference on Learning Representations

\bibitem[{{Yang} {et~al.}(2016){Yang}, {Jiang}, {Yang}, {Yu}, \&
  {Xu}}]{yang2016rapid}
{Yang}, B., {Jiang}, Y., {Yang}, J., {Yu}, S., \& {Xu}, Z. 2016, \apj, 816, 41,
  \dodoi{10.3847/0004-637X/816/1/41}

\bibitem[{{Zharkova} \& {Schetinin}(2005)}]{zharkova2005filament}
{Zharkova}, V.~V., \& {Schetinin}, V. 2005, \solphys, 228, 137,
  \dodoi{10.1007/s11207-005-5622-1}

\bibitem[{{Zhu} {et~al.}(2019){Zhu}, {Lin}, {Wang}, {Liu}, \&
  {Yang}}]{zhu2019solar}
{Zhu}, G., {Lin}, G., {Wang}, D., {Liu}, S., \& {Yang}, X. 2019, \solphys, 294,
  117, \dodoi{10.1007/s11207-019-1517-4}

\bibitem[{{Zhu} {et~al.}(2025){Zhu}, {Lin}, {Yang}, \& {Zeng}}]{zhu2025flat}
{Zhu}, G., {Lin}, G., {Yang}, X., \& {Zeng}, C. 2025, \apj, 980, 176,
  \dodoi{10.3847/1538-4357/adadff}

\bibitem[{{Zou} {et~al.}(2019){Zou}, {Jiang}, {Wei}, {Zuo}, \&
  {Wang}}]{zou2019statistical}
{Zou}, P., {Jiang}, C., {Wei}, F., {Zuo}, P., \& {Wang}, Y. 2019, \apj, 884,
  157, \dodoi{10.3847/1538-4357/ab4355}

\end{thebibliography}
\bibliographystyle{aasjournal}

\end{document}